\documentclass[10pt,letterpaper]{article}

\usepackage{cogsci}

\cogscifinalcopy %

\usepackage{cmap}
\usepackage[T1]{fontenc}
\usepackage[american]{babel}
\usepackage{csquotes}
\usepackage{newtxtext,newtxmath}  %

\usepackage[
  backend=biber,
  style=apa,
  natbib=true,
  annotation=false,
]{biblatex}
\usepackage{float} %

\usepackage[hidelinks]{hyperref}
\usepackage{cleveref}

\usepackage{graphicx}

\title{Learning to Trust: How Humans Mentally Recalibrate AI Confidence Signals}

\author[1]{\mbox{ZhaoBin Li (zhaobin.li@uci.edu)}}
\author[1]{\mbox{Mark Steyvers (mark.steyvers@uci.edu)}}
\affil[1]{Department of Cognitive Sciences, University of California, Irvine}

\begin{document}

\maketitle

\begin{abstract}
Productive human-AI collaboration requires appropriate reliance, yet contemporary AI systems are often miscalibrated, exhibiting systematic overconfidence or underconfidence. We investigate whether humans can learn to mentally recalibrate AI confidence signals through repeated experience. In a behavioral experiment ($N = 200$), participants predicted the AI's correctness across four AI calibration conditions: standard, overconfidence, underconfidence, and a counterintuitive ``reverse confidence'' mapping. Results demonstrate robust learning across all conditions, with participants significantly improving their accuracy, discrimination, and calibration alignment over 50 trials. We present a computational model utilizing a linear-in-log-odds (LLO) transformation and a Rescorla-Wagner learning rule to explain these dynamics. The model reveals that humans adapt by updating their baseline trust and confidence sensitivity, using asymmetric learning rates to prioritize the most informative errors. While humans can compensate for monotonic miscalibration, we identify a significant boundary in the reverse confidence scenario, where a substantial proportion of participants struggled to override initial inductive biases. These findings provide a mechanistic account of how humans adapt their trust in AI confidence signals through experience.

\end{abstract}

\section{Introduction}

From everyday scenarios like vacation planning to critical settings like medical triage, humans are increasingly relying on artificial intelligence to aid decision-making \citep{Green2019-hd, Agarwal2023-ub}.  The key to productive human-AI collaboration is knowing when to accept or reject the AI's recommendations—a phenomenon known as appropriate reliance \citep{Lee2004-tm, Tejeda2022-wv}. To support this, AI systems often provide confidence scores to help users gauge the reliability of a given suggestion on a case-by-case basis \citep{Zhang2020-yf}. Consequently, understanding how humans interpret and calibrate their trust in these uncertainty signals is essential to optimizing human-AI collaboration.

Prior research suggests humans often use AI confidence scores as direct indicators of reliability, trusting predictions when confidence is high and rejecting them when low \citep{Zhang2020-yf, Rechkemmer2022-ah}. Critically, AI systems frequently exhibit systematic miscalibration—expressing overconfidence or underconfidence relative to their actual accuracy. Contemporary AI systems, including neural networks and large language models, are particularly prone to overconfidence \citep{Guo2017-km, Jiang2020-so, Wang2023-oe}. This mismatch between reported and actual reliability causes humans to over- or under-rely on AI advice \citep{Ma2024-zv}, degrading both task performance and possibly trust in AI systems \citep{Li2025-zn}.

However, in real-world settings, humans are not static observers but engage with AI repeatedly over time, creating opportunities to improve their collaboration \citep{Bansal2019-kl, Tejeda2022-wv}. This raises an important question about experiential learning: can people learn to mentally calibrate AI confidence scores, adapting their reliance strategies to compensate for systematic biases in uncertainty signals?

This question bears directly on both the cognitive science of learning and the practical design of AI systems. From a cognitive perspective, mentally calibrating AI confidence scores poses a nontrivial learning challenge: humans need to aggregate prediction errors across repeated interactions, gradually updating an internal mapping between the AI's reported confidence and its actual probability of being correct, and then subsequently adjusting their reliance strategies in accordance based on their evolving mental model. This adaptive process naturally suggests the involvement of reinforcement-learning mechanisms, yet existing research has not directly investigated this topic.

The practical implications are equally significant: if humans cannot easily compensate for AI miscalibration through experience, then system designers need to prioritize calibration—potentially at the expense of other important metrics like accuracy \citep{Minderer2021-mz} and fairness \citep{Kleinberg2016-hq}, or requiring increased computational and data complexity \citep{Wang2023-oe, Huang2020-bh}. Conversely, if humans can learn to recalibrate AI confidence, then designers may deploy AI systems with greater miscalibration without sacrificing downstream performance, shifting design priorities toward supporting human–AI collaboration rather than purely AI calibration.

We investigate this question using a controlled behavioral experiment and computational modeling. In the experiment, participants learned over repeated interactions to predict whether an AI's response was correct based solely on its reported confidence. By manipulating the AI's calibration—including over‑confidence, under‑confidence, and a challenging reverse‑confidence condition in which higher AI confidence signaled lower accuracy—we show that participants exhibit robust learning across all conditions, steadily improving their accuracy, discrimination, and calibration alignment over trials. To explain the mechanisms underlying this adaptation, we develop a computational model demonstrating that participants mentally recalibrate AI confidence by updating their baseline trust and confidence sensitivity. Together, these results provide a principled account of human adaptation to unreliable information sources.

\begin{figure}[htbp]
    \centering
    \includegraphics[width=1\linewidth]{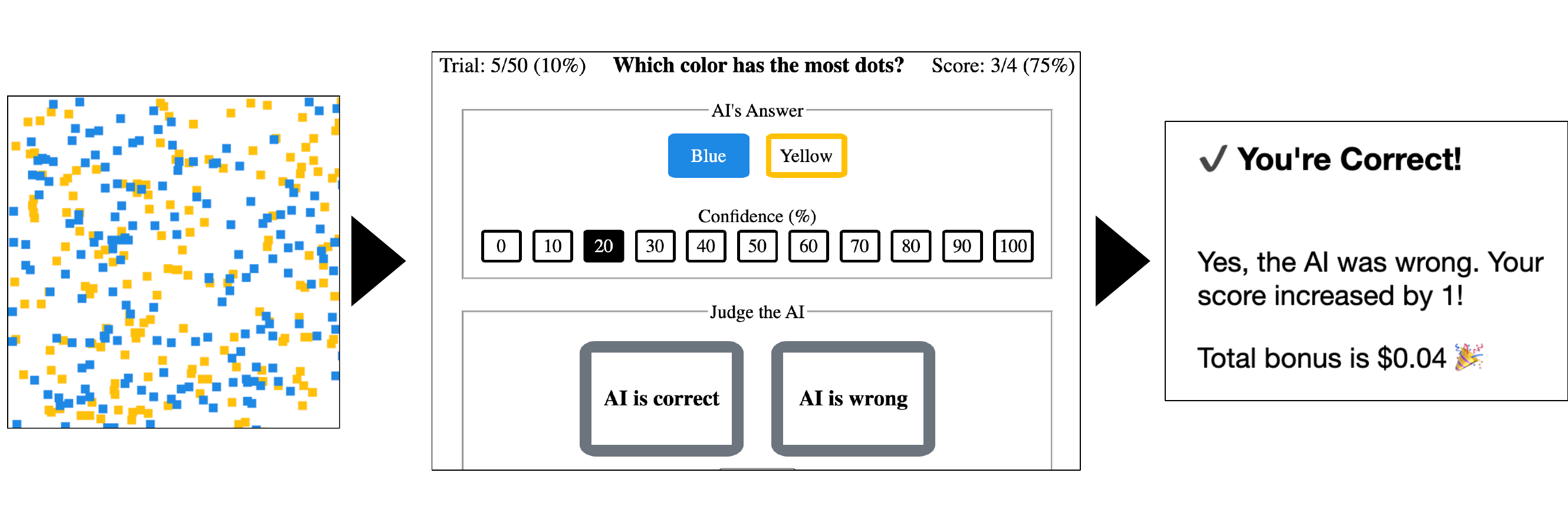}
    \caption{Experimental trial structure: participants viewed a 1-second colored dots animation, and then received the AI's prediction on which color has the most dots and confidence score rounded to the nearest 10\%. Participants then judge whether the AI was correct or wrong, and receive immediate feedback on their accuracy.}
    \label{fig:app}
\end{figure}

\section{Experiment}
\subsection{Methods}

To investigate whether humans can learn to mentally calibrate AI confidence signals through experience, we conducted an online behavioral experiment. Participants were tasked with learning to predict the correctness of an AI's response based solely on its reported confidence level. To simulate a realistic AI-assisted decision-making environment, we employed a cover story where participants performed a visual task and verified the AI's assessment of it, adapting established paradigms in human-AI collaboration \citep{Tejeda2022-wv, Liang2022-cn}.

\subsubsection{Participants} We recruited 200 participants via Prolific ($M_{\text{age}} = 43$, $SD_{\text{age}} = 13$; 55\% women, 45\% men). The study protocol was approved by our university's Institutional Review Board (IRB). The median time to complete the experiment was approximately 9 minutes. Participants received a base compensation of \$1.25 and were incentivized with an additional \$0.01 bonus for every correct answer, up to a maximum of \$0.50.

\subsubsection{Materials and Procedure}

The online experiment was implemented using JavaScript and the steps are illustrated in Figure 1. After providing informed consent and completing an onboarding tutorial and practice trial, participants entered the main experimental phase consisting of 50 trials.

On each trial, participants viewed a 1-second animation of moving dots and received the AI's prediction (which color had the most dots) along with its confidence score (0--100\%, rounded to nearest 10\%). Participants then judged whether the AI was ``Correct'' or ``Wrong.'' In reality, the dot counts were identical, so the only way to succeed was by learning the AI's calibration pattern. Participants were not told the AI's underlying accuracy or confidence distribution. Immediate feedback was provided after each judgment, showing whether the participant was correct, the change in their score, and their accumulated bonus.

\begin{figure}[htbp]
    \centering
    \includegraphics[width=1\linewidth]{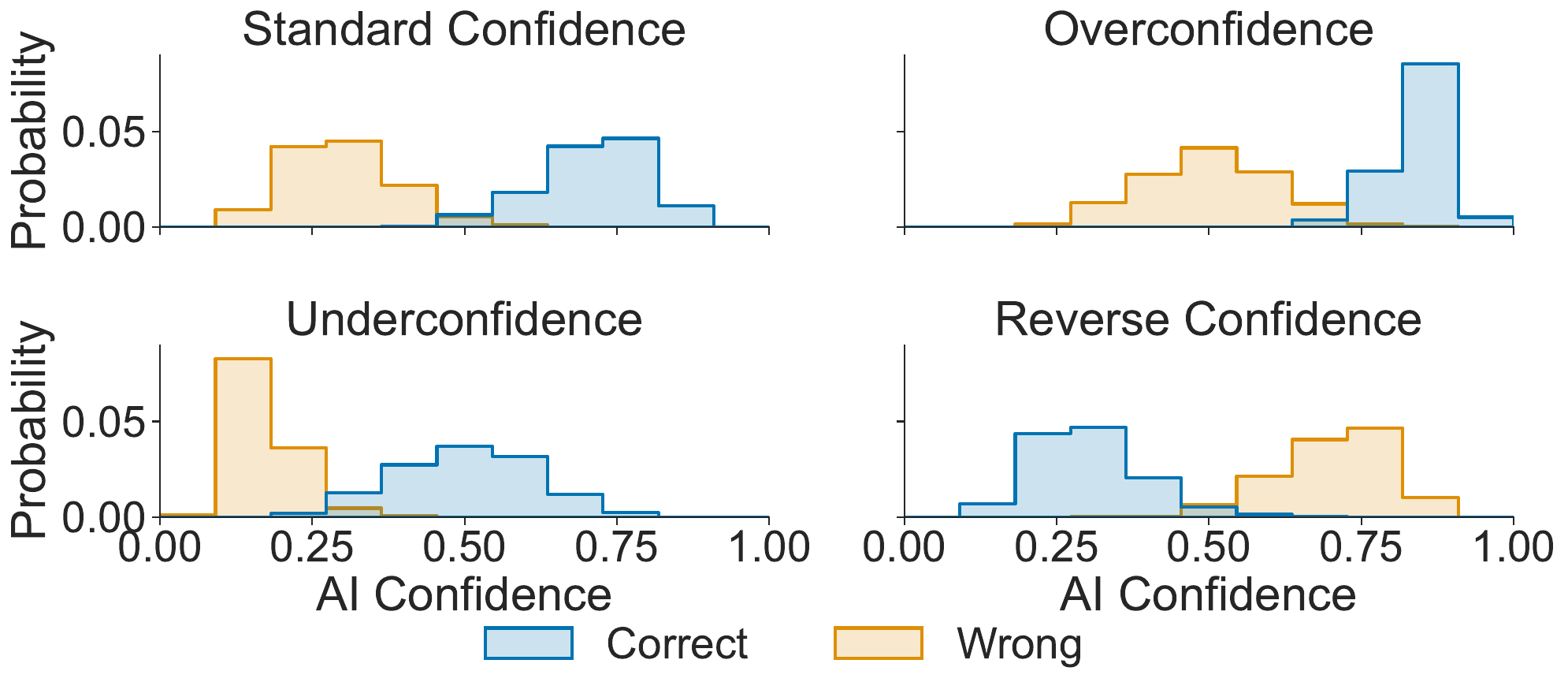}
    \caption{Probability densities of AI confidence distributions across the four conditions for correct and wrong decisions. The probabilities are rounded to the nearest 10\% as shown to the experimental interface and normalized across both distributions.}
    \label{fig:con}
\end{figure}

\subsubsection{Experimental Conditions} We utilized a between-subjects design in which participants were randomly assigned to one of four AI calibration conditions: standard confidence, overconfidence, underconfidence, and reverse confidence. In all conditions, the AI's overall accuracy was set at 50\% to ensure that any performance above chance resulted from genuine learning rather than simply always judging the AI as correct or wrong.
In all four conditions, the confidence scores were generated using a signal detection model in which confidence values for correct and wrong decisions were drawn from logit-normal distributions ($\sigma = 0.5$) with means separated by 2 units on the logit scale to promote learning (an ideal observer could judge the AI's correctness with 97\% accuracy). The confidence distributions are plotted in \Cref{fig:con} and the parameters are listed below: 
\begin{enumerate}
    \item \textbf{Standard Confidence:} Means for correct and wrong decisions were 1 and $-1$, respectively; the optimal decision criterion was 0.5.
    \item \textbf{Overconfidence:} Means were shifted to 2 and 0, resulting in a higher optimal decision criterion of 0.75.
    \item \textbf{Underconfidence:} Means were shifted to 0 and $-2$, resulting in a lower optimal decision criterion of 0.25.
    \item \textbf{Reverse Confidence:} Means were inverted relative to the standard confidence condition ($-1$ for correct, 1 for wrong); the criterion remained at 0.5 but the AI was more likely correct when reporting low confidence.
\end{enumerate}
In the experiment, we pre-generated 10,000 confidence scores per condition and randomly sampled from them per trial.

\subsection{Results}

\subsubsection{Accuracy Improved Across Trials}

\begin{figure}[htbp]
    \centering
    \includegraphics[width=1\linewidth]{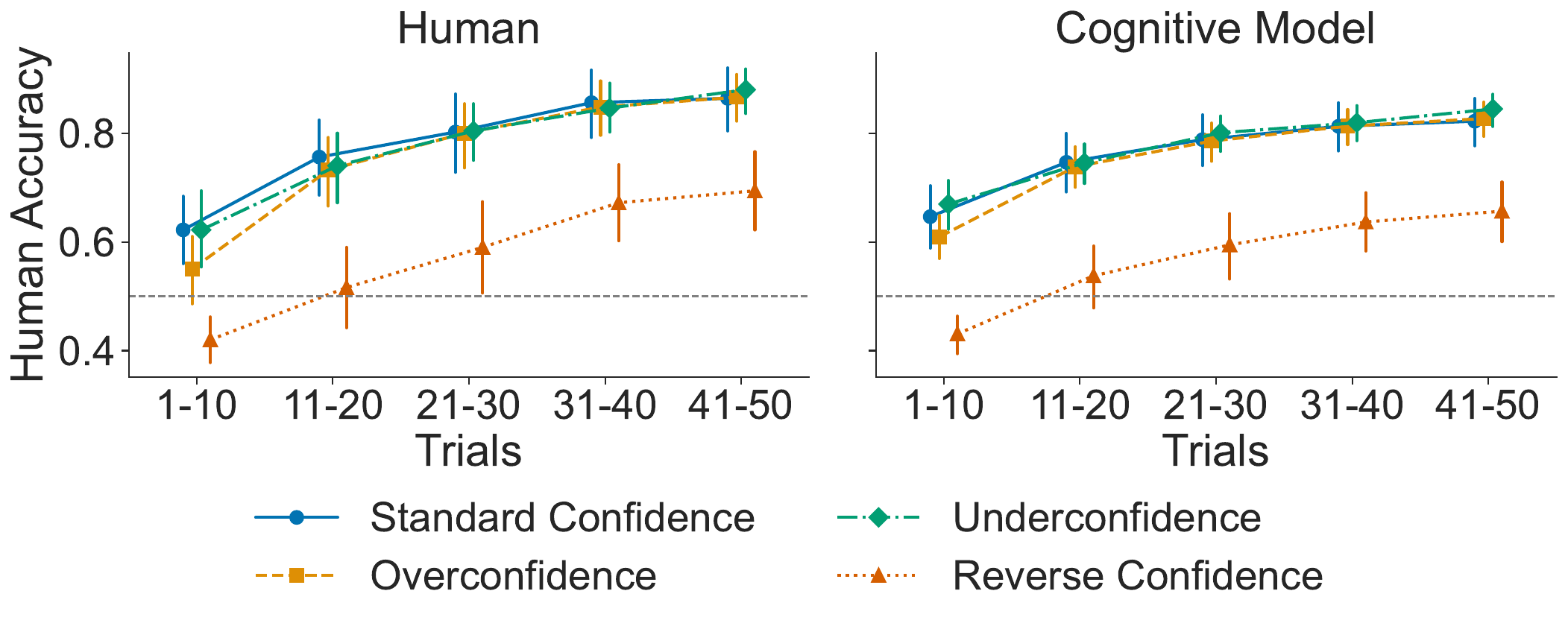}
    \caption{Accuracy improvement across trial blocks for human participants and cognitive model in all four AI calibration conditions. Error bars represent 95\% confidence intervals and the dashed line indicates 50\% chance accuracy.}
    \label{fig:acc}
\end{figure}

As shown in \Cref{fig:acc}, participants demonstrated clear learning across all four calibration conditions. Accuracy increased substantially over the 50 experimental trials, rising from early (trials 1--10) to late (trials 41--50) phases: 62\% to 86\% in standard confidence, 55\% to 87\% in overconfidence, 62\% to 88\% in underconfidence, and 42\% to 69\% in the reverse confidence condition. Even in the reverse confidence condition---designed to be maximally counterintuitive---participants improved markedly. Multilevel logistic regression predicting accuracy from trial number confirmed significant positive learning slopes across all conditions ($p < 0.001$,  $\beta$ from 0.049 to 0.064).

\begin{figure*}[htbp]
    \centering
    \includegraphics[width=1\linewidth]{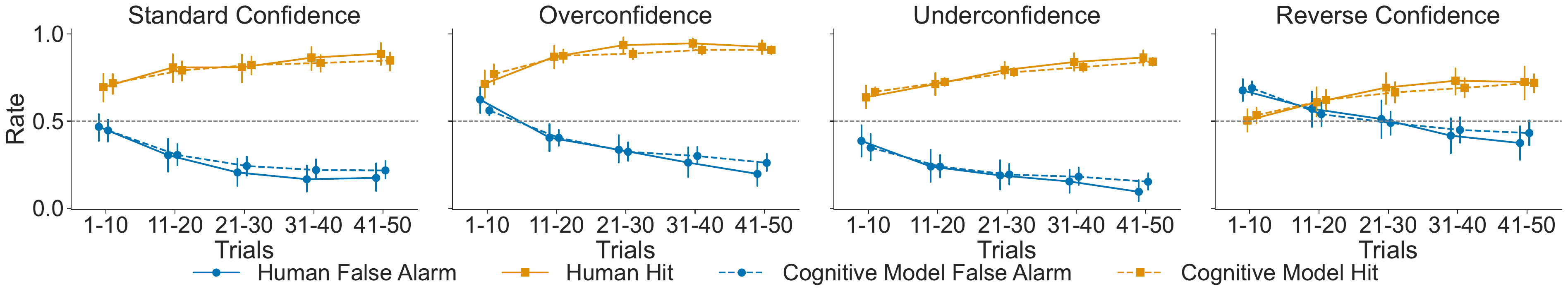}
    \caption{Hit rate increases and false alarm rate decreases across trial blocks for human participants and cognitive model in all four AI calibration conditions. Error bars represent 95\% confidence intervals and the dashed line indicates 50\% chance level.}
    \label{fig:bd}
\end{figure*}

\subsubsection{Hit and False Alarm Rates Improved Across Trials}

To better understand the mechanisms underlying these accuracy gains, we examined changes in hit rate (HR)---the proportion of trials where participants correctly identified the AI as ``correct''---and false alarm rate (FAR)---the proportion where participants mistakenly identified a wrong AI response as ``correct.''

As shown in \Cref{fig:bd}, HR and FAR improved significantly in all four conditions. In the standard confidence condition, HR increased from 72\% to 90\% and FAR decreased from 48\% to 18\%, demonstrating substantially improved discrimination. Participants in the overconfidence condition were initially misled by AI overconfidence, with early FAR at 62\%, yet successfully learned to correct for its bias, achieving 92\% HR and 19\% FAR in late trials. A similar pattern appeared in the underconfidence condition.

The most substantial learning occurred in the reverse confidence condition. Early on, participants' decisions aligned with the misleading confidence signals, resulting in 51\% HR and 68\% FAR. Nevertheless, by late trials, participants had partially inverted their decision rule, achieving 73\% HR and 35\% FAR.

Multilevel probit regression confirmed significant increases in HR ($\beta$ from 0.018 to 0.027, $p < 0.001$) and decreases in FAR ($\beta$ from $-0.027$ to $-0.036$, $p < 0.001$) across all conditions. Critically, sensitivity ($d'$) increased significantly in all conditions ($\beta$ from 0.045 to 0.062, $p < 0.001$), indicating that participants became more successful at distinguishing correct from wrong AI responses.

\subsubsection{Human Calibration to AI Improved Across Trials}

Participants also improved their calibration to the AI's confidence signals. We use the cognitive model in the next section to visualize the human calibration curves, because the pooled estimate is less noisy. To quantify the improvement in the raw data, we calculated the Expected Calibration Error (ECE) between the AI's accuracy and perceived AI accuracy per condition and phase \citep{Pakdaman-Naeini2015-pi}: $\text{ECE} = \sum_{m=1}^{M} \frac{|B_m|}{n} |\text{acc}_{\text{AI}}(B_m) - \text{acc}_{\text{human}}(B_m)|$, where $M = 11$ bins (confidence rounded to the nearest 0.1), $|B_m|$ is the number of trials in bin $m$, $n$ is the total number of trials, $\text{acc}_{\text{AI}}(B_m)$ is the AI's actual accuracy in that bin, and $\text{acc}_{\text{human}}(B_m)$ is the perceived AI accuracy (the proportion of trials where participants believed the AI was correct).

Paired $t$-tests confirmed significant ECE reductions from early to late trials ($p < 0.001$ for all conditions). Taken together, these results demonstrate that participants exhibited robust learning across conditions, not only improving in overall accuracy but also becoming more discriminating and substantially better calibrated to the AI---even when the AI's calibration was systematically distorted or inverted.

\section{Cognitive Modeling}
The behavioral results demonstrate clear learning but do not reveal the underlying cognitive mechanisms. Participants need to form beliefs about how AI confidence relates to its correctness, update these beliefs trial-by-trial based on feedback, and do so in ways that vary across individuals. We model participants' judgments using a linear-in-log-odds (LLO) transformation with two evolving parameters: an intercept capturing baseline trust in the AI, and a slope capturing confidence sensitivity. Both parameters are updated via a Rescorla–Wagner learning rule and estimated using Bayesian multilevel inference, enabling us to explain trial-by-trial dynamics, prior beliefs, and individual differences.

\subsection{Modeling Framework}
\subsubsection{Linear-in-Log-Odds Calibration} Consistent with the literature on human probability weighting \citep{Gonzalez1999-jq, Zhang2012-zx, Turner2014-go}, we assume that participants mentally recalibrate the AI's reported confidence using a linear-in-log-odds (LLO) transformation---their perceived AI accuracy $v_t$ on trial $t$ is linear with respect to the AI's reported confidence $c_t$ on a log odds scale:
\begin{equation*}
\log\left(\frac{v_t}{1-v_t}\right) = b_t + w_t \cdot \log\left(\frac{c_t}{1-c_t}\right)
\end{equation*}
This transformation maps confidence to perceived accuracy with an S-shaped curve (examples seen in \Cref{fig:cal}) and includes two interpretable parameters to capture shifts in participant beliefs. First, the intercept $b_t$ shifts the curve vertically and represents the participant's baseline propensity to trust the AI. We expect it to increase in the underconfidence condition and decrease in the overconfidence condition if participants learn to compensate for the AI's systematic bias.
 
Second, the slope $w_t$ captures human sensitivity to changes in the AI's confidence scores, with the sign of $w_t$ representing the perceived direction of correlation between the AI's confidence and accuracy. We expect $w_t$ to be positive a priori, reflecting the intuitive belief that higher AI confidence signals higher accuracy, and to become negative in the reverse confidence condition if participants learn that higher AI confidence signals lower accuracy.

\subsubsection{Rescorla–Wagner Learning Rule} We assume that both parameters—baseline trust $b_t$ and confidence sensitivity $w_t$—are updated trial by trial using a Rescorla–Wagner learning rule \citep{Rescorla1972-bt}, equivalent to the TD(0) update used in reinforcement learning \citep{Sutton2018-ls}. After every trial, participants observe whether the AI's prediction was correct ($g_t = 1$) or wrong ($g_t = 0$) and compute a prediction error $\delta_t = g_t - v_t$.

Learning rates for the two parameters $b$ and $w$ were kept separate and asymmetric for positive and negative errors, reflecting established patterns in human reward learning \citep{Katahira2018-gn, Gershman2015-ci}. As the results will show, the learning rates vary substantially across $b$ and $w$ and by the sign of the errors. Because the sign of the error corresponds to the AI's accuracy, the learning rates are indexed by $g_t$. To minimize the binary cross-entropy loss, the parameters are updated according to:
\begin{align*}
b_{t+1} &= b_t + \alpha_{b,g_t} \cdot \delta_t \\
w_{t+1} &= w_t + \alpha_{w,g_t} \cdot \delta_t \cdot \log\left(\frac{c_t}{1-c_t}\right)
\end{align*}
In total, six parameters are estimated per participant: the initial bias $b_0$, the initial weight $w_0$, and four learning rates ($\alpha_{b,g_t=1}$, $\alpha_{b,g_t=0}$, $\alpha_{w,g_t=1}$, and $\alpha_{w,g_t=0}$).

\subsubsection{Bayesian Multilevel Inference} Parameters were estimated using multilevel Bayesian modeling to improve precision and capture individual variability while obtaining group-level trends. On every trial $t$, participant $i$'s binary response $y_{t,i}$ (coded as 1 if the participant judged the AI correct, 0 otherwise) is modeled as a Bernoulli draw with parameter $v_{t,i}$: $$y_{t,i} \sim \text{Bernoulli}(v_{t,i}).$$
Because participants were randomly assigned to conditions, we applied an experiment-wide prior to their initial biases and weights: $b_{0,i} \sim N(\mu_{b_0}, \sigma_{b_0})$, $\mu_{b_0} \sim N(0, 1)$, and $\sigma_{b_0} \sim \text{Exponential}(1)$. This prior centers the initial bias at 50\% perceived accuracy, providing a 95\% prior interval of [12\%, 88\%] to allow for wide variability in baseline trust. An identical prior structure was applied to $w_0$.

Furthermore, given that learning rates are likely influenced by the specific AI condition $c$, we utilized separate condition-wide priors: $\log\left(\frac{\alpha_{b,g_t,i,c}}{1-\alpha_{b,g_t,i,c}}\right) \sim N(\mu_{\alpha_{b,g_t,c}}, \sigma_{\alpha_{b,g_t,c}})$, $\mu_{\alpha_{b,g_t,c}} \sim N(-1.5, 1.5)$, and $\sigma_{\alpha_{b,g_t,c}} \sim \text{Exponential}(1)$. The hyper-mean was set to center the learning rate around $\approx 0.18$. The log-odds transformation ensures that all $\alpha$ values remain within the $[0, 1]$ interval.

The model was implemented in Stan \citep{Carpenter2017-vy} using Markov Chain Monte Carlo (MCMC) with 4 chains and 2,000 samples per chain. Convergence was verified via trace plots and by ensuring $\hat{R} < 1.01$ and ESS $> 400$.

\subsection{Modeling Results}

\subsubsection{Cognitive model predicts behavioral trends}

The behavioral results demonstrate clear learning, but to understand the underlying mechanisms, we need a computational account. Our cognitive model captures the key behavioral patterns across all four conditions, including improvements in overall accuracy (\Cref{fig:acc}) and changes in hit and false alarm rates (\Cref{fig:bd}). On average, model predictions match human decisions on 75\% of trials (mean log-likelihood per trial $= -0.38$, McFadden's pseudo-$R^2 = 0.45$), substantially exceeding 50\% chance performance. This level of agreement indicates that a simple belief-mapping framework with trial-by-trial updating is sufficient to reproduce participants' aggregate reliance behavior.

\begin{figure*}[htbp]
    \centering
    \includegraphics[width=1\linewidth]{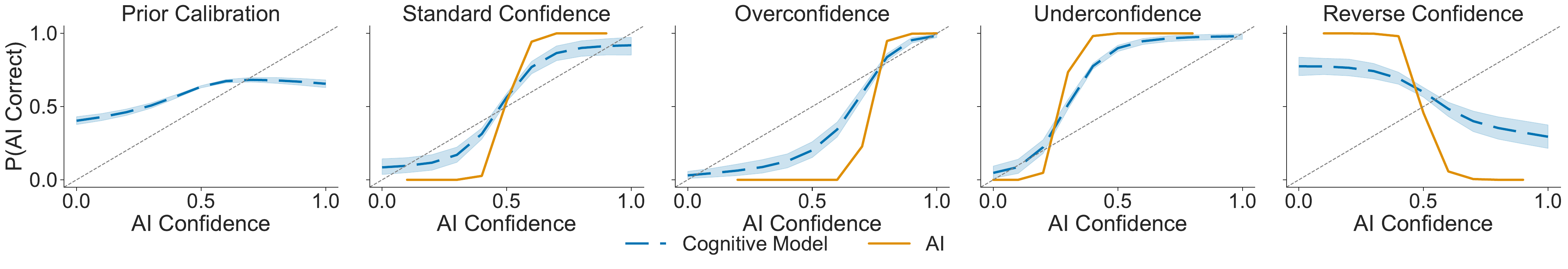}
    \caption{Calibration curves showing AI accuracy (orange) and human perceived AI accuracy (blue dashed) across conditions. Left panel shows prior calibration at trial 1 from experiment-wide priors. Subsequent panels show posterior calibration at trial 50 from participant-level posteriors per experimental condition. Black diagonal dashed line shows perfect calibration. Shaded regions show 95\% credible intervals.}
    \label{fig:cal}
\end{figure*}

\subsubsection{Changes in belief mapping of AI confidence and perceived accuracy over time}

The behavioral analysis established that calibration improves, but the noise in raw data prevents us from recovering participants' initial beliefs or their learned mappings. To address this, we use the cognitive model to infer how participants mentally translate AI confidence into perceived accuracy.

To estimate \textit{a priori} calibration, we sampled 1,000 values of the initial parameters ($b_0$ and $w_0$) from the experiment-wide priors and computed perceived AI accuracy across all confidence values. \Cref{fig:cal} (left panel) shows these inferred prior calibration curves. As expected, participants begin with positive confidence sensitivity ($M = 0.69$, $SD = 1.8$), believing that higher AI confidence signals higher accuracy. They also exhibit positive baseline trust ($M = 0.60$, $SD = 0.56$), initially believing that the AI is approximately 65\% accurate when it reports 50\% confidence.

To obtain posterior calibration at the end of learning, we used participant-level posterior means of the parameters at trial 50 ($b_{50}$ and $w_{50}$). The resulting curves in \Cref{fig:cal} (subsequent panels) reveal successful adaptation: learned calibration closely tracks the AI calibration in the standard, overconfidence, and underconfidence conditions. However, the reverse confidence condition shows only partial alignment, reflecting substantial individual differences we explore next.

\begin{figure}[htbp]
    \centering
    \includegraphics[width=1\linewidth]{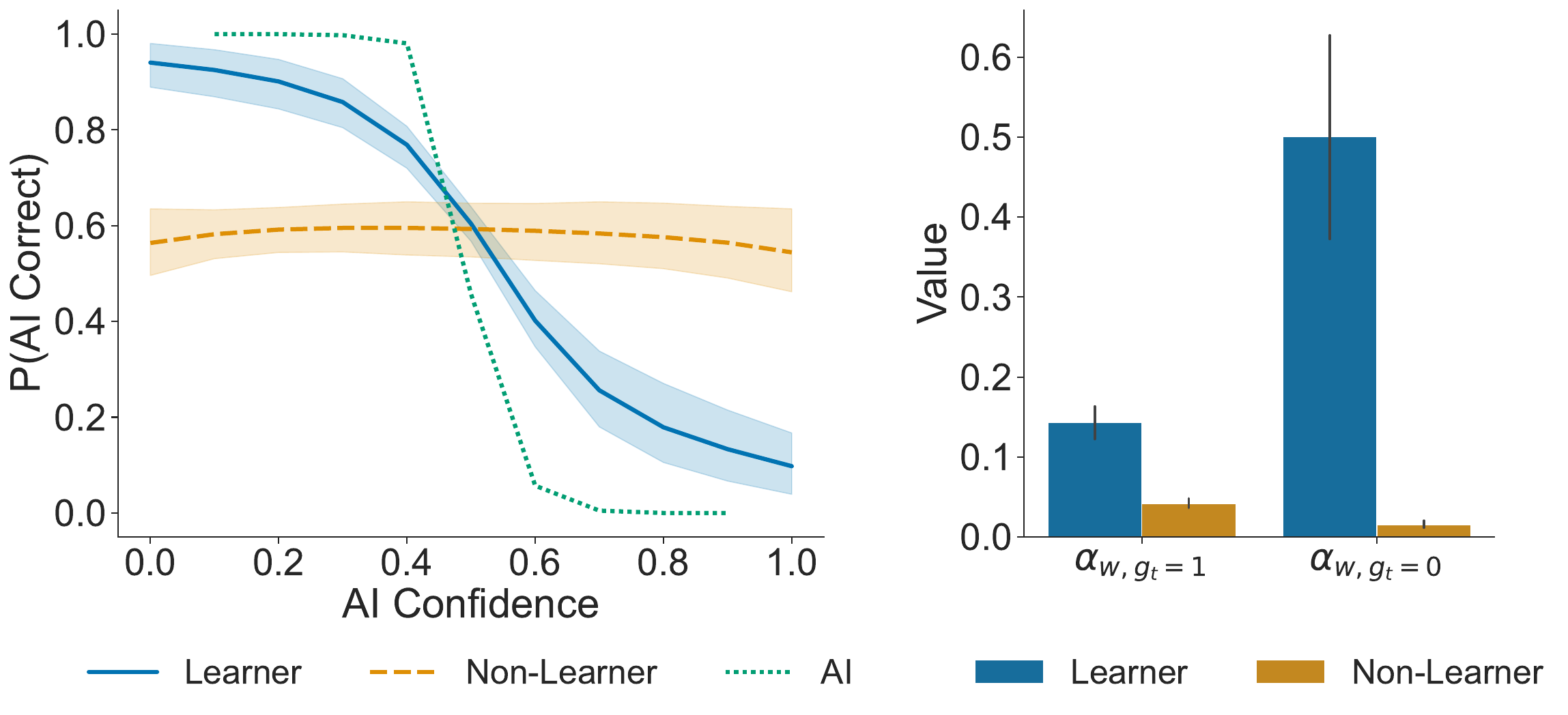}
    \caption{Individual differences in the reverse-confidence condition. \textbf{Left}: Calibration curves at trial 50 showing AI accuracy (dashed green) and human perceived AI accuracy for learners (blue) and non-learners (orange). \textbf{Right}: Learning rates for confidence sensitivity ($w$) when the AI was correct ($\alpha_{w,g_t=1}$) and wrong ($\alpha_{w,g_t=0}$). Shaded regions and error bars represent 95\% credible intervals.}
    \label{fig:rev_dif}
\end{figure}
 
\subsubsection{Explaining Individual Differences in Reverse Confidence}

The reverse confidence condition posed a uniquely counterintuitive challenge---participants had to learn that higher AI confidence actually signaled lower accuracy. While participants generally improved over time, a detailed analysis of individual accuracy revealed that a substantial proportion struggled to adapt. Categorizing participants by whether their late-stage accuracy (trials 31--50) exceeded 60\%, we found that 44\% in the reverse confidence condition were non-learners---a stark contrast to the other three conditions, where non-learners comprised less than 15\% of the sample.

\Cref{fig:rev_dif} (left panel) shows the estimated calibration curves at trial 50, separated by learner status. Learners eventually matched the negative slope of the AI's actual calibration (mean $w_{50} = -2.15$), successfully inverting their initial beliefs. In contrast, non-learners' calibration remained largely flat (mean $w_{50} = 0.01$), retaining a near-zero sensitivity to confidence signals. Critically, both groups began with similar positive priors ($w_0 > 0$), yet only learners managed to override their initial inductive bias.

The cognitive model reveals the mechanism behind this divergence. \Cref{fig:rev_dif} (right panel) shows that learners exhibited dramatically higher learning rates for confidence sensitivity ($w$) than non-learners. When the AI was correct, learners updated their sensitivity nearly 3$\times$ faster ($\alpha_{w,g_t=1} = 0.14$ vs. $0.04$); when the AI was wrong, this multiple expanded to 30$\times$ ($0.50$ vs. $0.015$). These results suggest that learners rapidly incorporated prediction errors to update their confidence sensitivity, while non-learners remained resistant to changing their initial mental calibration. Importantly, learning rates for baseline trust ($b$) were similar between groups ($p > 0.05$), confirming that the model successfully isolates the specific cognitive mechanism responsible for individual differences.

\begin{figure}[htbp]
    \centering
    \includegraphics[width=1\linewidth]{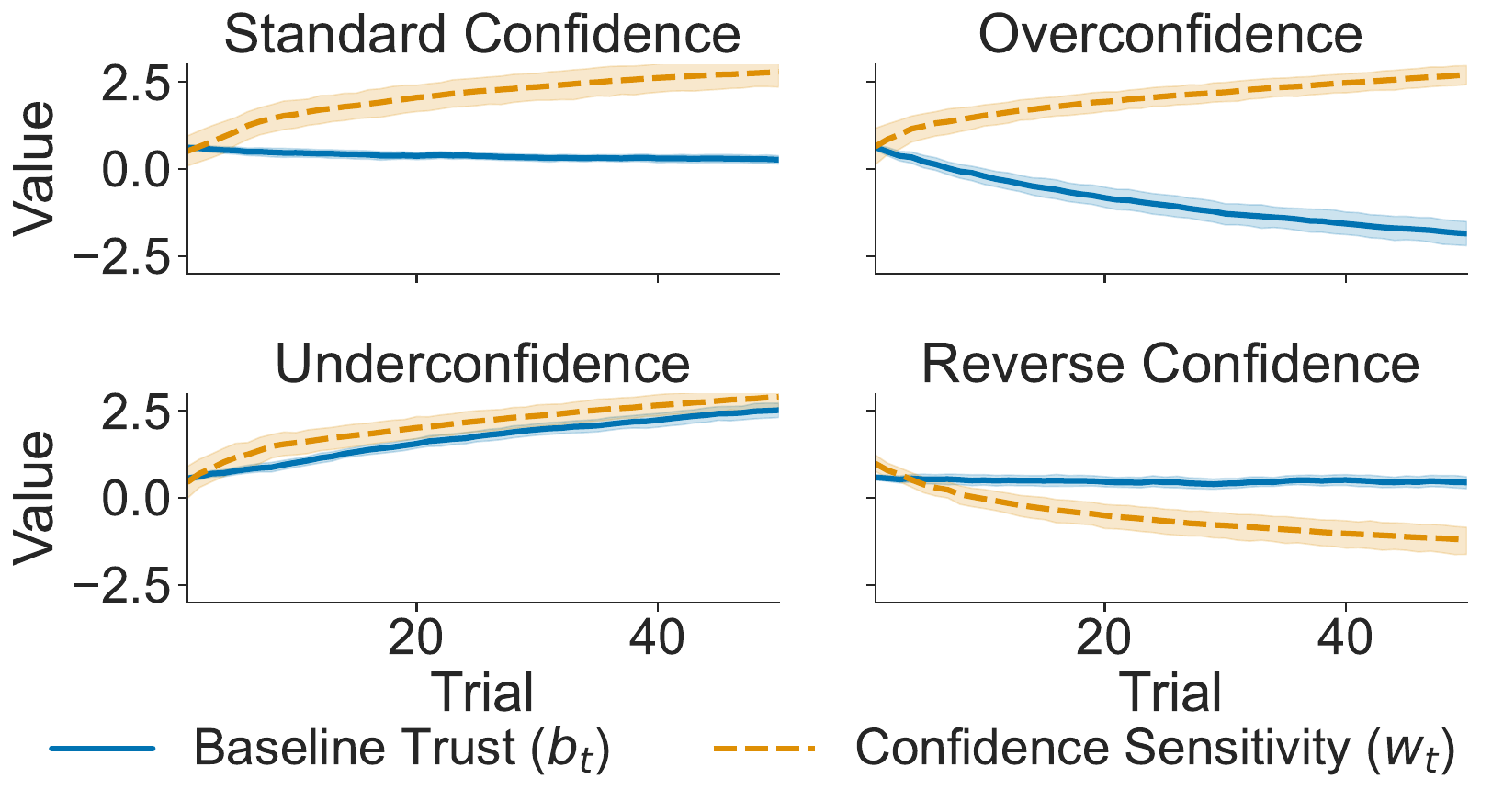}
    \caption{Changes in baseline trust $b_t$ (solid blue) and confidence sensitivity $w_t$ (dashed orange) across trials in all four conditions. Shaded regions represent 95\% credible intervals.}
    \label{fig:trace}
\end{figure}

\subsubsection{Participants adapt baseline trust and confidence sensitivity}

To understand how learning evolves, we examined changes in the model's two key parameters across trials (\Cref{fig:trace}). Confidence sensitivity ($w_t$) captures how participants weight AI confidence signals. In the standard, overconfidence, and underconfidence conditions, $w_t$ increased substantially---from 0.45--0.65 at trial 1 to 2.7--2.9 at trial 50---demonstrating that participants became more attuned to AI confidence as a diagnostic signal. In the reverse confidence condition, $w_t$ decreased from 0.59 to $-1.2$, indicating that participants successfully learned the inverse mapping where higher confidence signaled lower accuracy.

Baseline trust ($b_t$) represents participants' tendency to trust the AI, independent of its reported confidence. This parameter adapted to compensate for systematic AI biases. In the underconfidence condition, baseline trust increased substantially (from $0.57$ to $2.5$) as participants learned that the AI was more accurate than it claimed. Conversely, in the overconfidence condition, trust decreased markedly (from $0.62$ to $-1.9$), reflecting learned skepticism toward AI overconfidence. In the standard and reverse conditions, trust decreased slightly (from $0.61$ to $0.27$ and $0.6$ to $0.45$, respectively) to correct for initial overtrust. All changes were statistically significant ($p < 0.001$).

\subsubsection{Asymmetric learning rates prioritize the most informative errors}

A key insight from the model is that participants do not update their beliefs uniformly---instead, they exhibit asymmetric learning that prioritizes the most diagnostic feedback. These asymmetries vary by condition, reflecting which outcomes are most informative for learning each AI's calibration pattern.

In the overconfidence condition, the most informative signal is when a highly confident AI makes an error. Accordingly, updates for baseline trust are substantially higher when the AI is wrong ($\alpha_{b,g_t=0} = 0.46$) than when it is correct ($\alpha_{b,g_t=1} = 0.29$), indicating participants rapidly lose trust when observing confident mistakes. Conversely, updates for confidence sensitivity are higher when the AI is correct ($\alpha_{w,g_t=1} = 0.55$ vs. $\alpha_{w,g_t=0} = 0.05$)---when an extremely confident AI proves correct, participants learn to give significantly more weight to its confidence signals.

As expected, these patterns reverse in the underconfidence condition, where the opposite pattern emerges. Baseline trust updates more when the AI is correct ($\alpha_{b,g_t=1} = 0.51$ vs. $\alpha_{b,g_t=0} = 0.14$), while confidence sensitivity updates more strongly when the AI is wrong ($\alpha_{w,g_t=0} = 0.49$ vs. $\alpha_{w,g_t=1} = 0.04$). The standard and reverse confidence conditions show more symmetric learning rates, likely because both correct and incorrect outcomes provide comparable information about the AI's calibration. Together, these results demonstrate that human learning strategically prioritizes feedback that best reveals the AI's underlying calibration structure.

\section{Discussion}
Our results demonstrate that humans can learn to mentally recalibrate AI confidence signals through repeated experience. Across four calibration conditions, participants substantially improved their accuracy, discrimination, and calibration alignment over only 50 trials. Our computational model reveals that this adaptation occurs through trial-by-trial updates to baseline trust and confidence sensitivity, with asymmetric learning rates that prioritize the most informative errors.

While prior research has examined how people calibrate their own confidence in human-AI contexts \citep{Chong2022-yq, Ma2024-zv} or recalibrate human advisers \citep{Pescetelli2019-cf, Stanciu2022-pn}, we demonstrate that humans adaptively recalibrate confidence signals generated by AI systems. This extends established frameworks of metacognition \citep{Fleming2012-uk, Koriat2012-rh} to human-AI collaboration and provides a computational account of how appropriate reliance develops through experience \citep{Lee2004-tm, Muir1994-mv}. The success of our reinforcement-learning model suggests that this calibration process engages error-based learning mechanisms \citep{Daw2006-ln, Sutton2018-ls}, consistent with recent findings that humans learn their own confidence through prediction errors \citep{Le-Denmat2026-ld}.

Our model reveals that baseline trust and confidence sensitivity are key components of human mental models of AI calibration \citep{Bansal2019-kl, Kelly2023-sj}. By decomposing trust into the two interpretable parameters, we provide a clear way to characterize how users miscalibrate AI confidence signals. This decomposition helps identify whether errors arise from overall over- or under-trust versus misinterpretation of confidence levels.

There are several limitations worth noting about our empirical approach. First, participants completed only 50 trials in a single session, and longer-term dynamics including memory consolidation, decay, or transfer remain unexplored. Second, immediate feedback was provided, but learning would likely be more challenging with delayed or absent ground truth. Third, participants in the task did not have the opportunity to make independent predictions. They could only learn from the AI's confidence and feedback, but real-world collaboration often allows some degree of independent verification. Future research should address these limitations.

\section{Conclusion}
We demonstrate that humans possess substantial capacity to adapt to AI confidence signals through reinforcement learning mechanisms. By showing that humans can compensate for monotonic miscalibration but struggle with inverse mappings, we identify both the potential and boundaries of human adaptability in collaborative AI systems. These findings contribute to theories of metacognition and social learning in human-AI collaboration while providing insights for designing systems that support appropriate reliance.

\section{}

\printbibliography

\end{document}